\documentclass[12pt]{article}

\usepackage{geometry}
\usepackage{fullpage}
\usepackage{graphicx}
\usepackage{pdflscape}
\usepackage{multirow}

\usepackage{float}
\usepackage{dcolumn}
\usepackage{booktabs,calc}
\usepackage{array}
\usepackage{paralist}
\usepackage{verbatim}
\usepackage{subfig}
\usepackage{setspace}
\usepackage{natbib}
\usepackage{amsmath} 
\usepackage{amssymb}
\usepackage{enumerate}
\usepackage[linktocpage=true, colorlinks=true, linkcolor=black, citecolor=black, urlcolor=black]{hyperref}
\usepackage{csquotes}
\usepackage{capt-of}
\usepackage{hyperref}
\usepackage{framed}
\usepackage{authblk}

\begin{document}

\title{Intra-Household Management of Resources: Evidence from Malawi}

\author{Anna Josephson}

\affil{\small{Department of Agricultural and Resource Economics, University of Arizona \\ \href{mailto:aljosephson@arizona.edu}{aljosephson@arizona.edu}}}

\date{\small{\today}} 

\maketitle

\normalfont

\begin{abstract}
\noindent In models of intra-household resource allocation, the earnings from joint work between two or more household members are often omitted. I test assumptions about complete pooling of resources within a household, by accounting for income earned jointly by multiple household members, in addition to income earned individually by men and women. Applied in the case of Malawi, I find that by explicitly including intra-household collaboration, I find evidence of partial income pooling and partial insurance within the household, specifically for expenditures on food. Importantly, including joint income reveals dynamics between household members, as well as opportunities and vulnerabilities which may previously be obfuscated in simpler, binary specifications. Contrasting with previous studies and empirical practice, my findings suggest that understanding detailed intra-household interactions and their outcomes on household behavior have important consequences for household resource allocation and decision making. 
\end{abstract}

\hfill

{\small \noindent\emph{JEL Classification}: O12, D13, J16  
		}

\newpage 
\doublespacing

\section{Introduction}

Household decision making traditionally assumes that all income is earned by a single household member, who is then responsible for those earnings in intra-household resource allocation \citep{Chiappori92, Browningetal94, BrowningandChiappori98}. Much of the existing literature addresses this issue by accounting for interactions between these individuals within the household, focusing on the interactions between men and women \citep{Udry96, DufloandUdry04, Basu06, Bobonis09}. But for many households, income earned jointly, by two or more household members, constitutes a large proportion of earned income. Work from Africa shows that between 12 percent and 70 percent of household plots are jointly managed, resulting in a substantial amount of shared income \citep{Kilicetal15, Slavchevska15, Harmonyetal2021}. The failure to consider these joint income-earning relationships has likely led to inaccuracies, misestimation, and misunderstanding of household dynamics in studies of intra-household income and resource allocation. In this paper, I relax assumptions about income earning, including income earned jointly between two or more household members, so as to better understand how households internally cope with and manage transitory shocks. This elucidates intra-household functioning and behavior, helping us to better understand how households operate during times of stress. 

In many households, income earned jointly by two or more household members, constitutes a large proportion of earned income. This is particularly true in Malawi, the context for this paper. Malawi is a nation in which intra-household collaboration is common, particularly in agriculture \citep{Kilicetal15, McCarthyandKilic17, DossandQuisumbing20}. To evaluate these collaborative interactions, I empirically test assumptions about household resource allocations. Specifically, I explore disparities in expenditure by different income earners, after the experience of a short term shock. My analysis focuses on three types of income earned in Malawi: that earned exclusively by a man, that earned exclusively by a woman, and that earned jointly by two or more people. I use these groups to test a restriction of the collective household model: that income from all of these income types is always pooled. The central observation underlying my methodological approach and empirical estimation is that if households completely pool income, then households members with different income types fully insure one another against short term fluctuations in their incomes. Thus, non-persistent income shocks will not result in changes in allocation of resources within the household. 

I reject the hypothesis of complete income pooling within the household. This means that household members do not fully insure one another against short term fluctuations in income. However, I find evidence that household members partially pool income for expenditures on food. This behavior results in partial pooling and partial insurance for the household on the most important purchase for many households, as well as the largest single expenditure item in household budgets. So, although households do not completely pool income, they exhibit a strategic pooling behavior for an essential expenditure. Further, my results reveal trends in income stability among household members and elucidate how households account for changes in income - and in particular how household members adjust their incomes when there are shocks to different household income types. In particular, the results reveal how changes in income from women shifts expenditure on a variety of goods. When including joint income, the impacts of changes in women's income on expenditure are evident, relative to men's and joint incomes. These findings elucidate intra-household dynamics further than previous studies. Overall, while I observe similar aggregate trends when including or excluding joint income, its inclusion demonstrates the inequities felt among household members, due to short term shocks. 

Further, I examine the role of household structure on intra-household resource allocation. I consider differences between male- and female-headed households, as well as households with patrilineal and matrilineal lineage. I find that both male- and female-headed households, as well as patrilineal and matrilineal lineage households, exhibit similar same behavior as the pooled sample. This suggests that regardless of household makeup, extending the standard specification to include joint income elucidates structural decisions as well as opportunities and vulnerabilities therein. This, interestingly, differs from previous literature which suggests there are likely to be significant differences based on lineage and household headship. Some of this may be due to the unique context that is Malawi: more than two thirds of households report a matrilineal heritage, but still nearly 80 percent of households are male-headed. This configuration may explain some of the disparities from previous literature, at least with respect to household structures. 

To test the assumption that all household income is pooled, I use a model developed by \cite{DufloandUdry04}, which allows me to explore the disparities in expenditure by different income earners resulting from exogenous variation in rainfall. The central observation underlying the methodological approach and empirical estimation is that if households completely pool income, then household members fully insure one another against short term fluctuations in income. If this is the case, then non-persistent income variations will not result in changes to the allocation of resources within the household.

The identification strategy relies on observing the impact of exogenous variation in rainfall on income and expenditure. I examine short term variations in rainfall, which are covariate at the household level. All members of a household experience the same rainfall, but the pattern of rainfall may have different impacts on individual household members, due to discrepancies in input use, cultivated crops, and plot quality. As previous work has shown, women and men tend to cultivate different portfolios of crops and do so on plots of different qualities and different quantities and qualities of inputs \citep{Udryetal95, Udry96, deBrauw15, ChenCollins14, Kilicetal15, Osen15, Slavchevska15, Doss18}. Thus, rainfall may have different impacts on the income of different household members and household labor allocation may be disparately impacted, by gender, after the experience of a shock \citep{BrancoFeres20, JosephsonandShively21}. In a household with complete insurance from the pooling of resources of all household earners, the different impacts of rainfall on different income earners, conditional on total expenditure, should not translate into differences in the allocation of a particular expenditure to different purposes within the household.

The results of this study suggests that by omitting joint relationships in the analysis of household behavior, research has failed to account for important dynamics in household analyses. Individual resource allocation decisions are integral to overall household income and spending behaviors. Interactions between these individuals are an essential part of understanding household behavior. This study extends our understanding of these important dynamics by including jointly earned income in the analysis. This paper is among the first to explicitly incorporate jointly earned income with individual income into the analysis of household income pooling.\footnote{There is a growing body of literature, centered in West Africa, which explores the concept of shared plots, which are generally managed by men. Although this literature highlights income which, in principle, may be classified as joint, it is not identified in this way \citep{DufloandUdry04, KaziangaandWahhaj13}.} While other areas have included aspects of joint management, this consideration has been omitted from studies of gender-specific household resource allocation.\footnote{\cite{Bobonis09}, \cite{Kilicetal15}, \cite{Osen15}, \cite{Slavchevska15}, \cite{McCarthyandKilic17} all include a discussion of joint collaboration but focus on production and shared household and agricultural responsibilities. Similarly, the idea of joint household appears in \cite{Bernardetal20} and \cite{Ambleretal22}, though related to survey design and the importance of household structure in that process.} The evidence presented in this study supports the validity and empirical relevance of including income earned jointly in intra-household analyses. It further contributes to a growing literature on how detailed understanding of intra-household dynamics and intra-household estimation may better help inform effective and targeted policy \citep{DSouzaandTandon20, Brownetal21, Dizonetal22}.

\section{Country Context and Data}\label{sec:e1data}

\subsection{Country Context}

The context for this study is Malawi. Malawi is an ideal location to examine intra-household relationships as collaboration in agricultural household activities is common in the country \citep{Kilicetal15, McCarthyandKilic17}. Inheritance is matrilineal, with matrilineal settlement practices \citep{Walther16}. However, traditional behavior gender behavior is also standard, especially in rural areas, and drives different household practices and outcomes: while more than two thirds of households report a matrilineal heritage, but still nearly 80 percent of households are male-headed. These circumstances lead to differential outcomes, in literacy, income, labor market participation, and poverty, for men and women. Poverty in the country is widespread and poverty rates differ between female- and male-headed households. While poverty rates for male-headed households have fallen, rates for female-headed households are stagnant \citep{Kilicetal15}. Further, \cite{Kilicetal15} find that, on average, female-managed plots are 25 percent less productive than male-managed plots. This discrepancy in productivity may drive some of the observed gendered poverty gap.

Crop choice and cropping behavior may also drive some of this observed gap. As \cite{Eicher90} report, cash crops and agricultural technologies in Malawi were historically introduced only to men. Thus, men developed a cash-crop, export-oriented agricultural practice while women continued cultivation of subsistence crops. This trend persists, as \cite{Mathiassenetal07} find that men are more likely to grow hybrid maize and tobacco, while women are more likely to grow local maize varieties. We observe similar trends, reported in Table~\ref{tab:e1cropsmf}: while maize is grown at similar rates, regardless of if the plot is managed by men, women, or managed jointly, men's and jointly managed plots are much more likely to grow tobacco, while women's plots are more likely to grow beans, including soybeans and pigeonpea. 

\subsection{Data}\label{sec:data}

Data for this study comes from two rounds of the Malawi Integrated Household Panel Survey (IHPS) which was implemented by the National Statistical Office under the Living Standard Measurement Survey - Integrated Household Surveys on Agriculture (LSMS-ISA) program with the World Bank. The sample was designed to be representative at the national, urban/rural, and regional levels. Efforts were made to track resettled and split off households. In 2013, the IHPS succeeded in tracking 3,246 households across 204 enumeration areas, which were surveyed in the previous round in 2010. Attrition is low, at 3.8 percent for households and 7.4 percent for individuals \citep{McCarthyandKilic17}. The data set contains information on demographics, expenditure, and agriculture, as well as household-specific rainfall measurements. 

The empirical analysis uses a balanced panel from households whose data was collected in both 2010 and 2013. Due to interest in income earned by agricultural households, I omit from the analysis households which did not report income from crop sales. Each household included in the sample has at least one household member who earned some income from the sale of crops. In 2010, 1,198 individuals (46 percent of individuals in the first round) reported selling some crops. In 2013, 1,720 individuals (52 percent of individuals in the second round) reported selling some crops. In total, 850 households appear in both years, yielding a total of 1,700 household observations. Because of the focus on a subset of the population, one cannot consider the sample nationally representative of Malawi. However, it is representative of individuals and households who participate in crop markets in Malawi. Further, rising sales and expanding markets across Africa suggest that these trends will be instructive, as more households begin to participate in crop and output markets \citep{Timmer17}.

Summary statistics for the variables used in our specifications are presented in Table~\ref{tab:sumstat}. I use three types of data. First, I use income data, which are classified according to two types of specifications: first a ``traditional'' specification, which includes only income from men and women, and next an ``extended'' specification, which includes joint income. Next, I use expenditure data, for a series of ten different types of household expenditures, including food; alcohol and tobacco; clothing; recreation; education; healthcare; housing and utilities; transportation; communication; and hotels and restaurants. Finally, I use rainfall data, in order to estimate the short term shocks experienced by household members. Each of these are discussed in more detail below. 

First, I consider the various types of income. In addition to details on the crops sold by a household, the LSMS surveys designate the household member who is responsible for decisions about the income earned from the sale of each crop. Respondents are asked to report the primary income manager and, as appropriate, a secondary income manager. The data are used to classify income managers. First, I create the ``traditional'' measure: only male and female income included (Table~\ref{tab:sumstat}, Panel 1 ``Omitted Joint Income''). In this measure, I consider only the primary manager and designate male and female income based on the gender of the individual reported. Next, I create the ``extended'' measure: with male, female, and joint income included (Table~\ref{tab:sumstat}, Panel 1 ``Joint Income''). To do this, I designate income as joint if a secondary income manager is identified. The analysis is indifferent about whether joint funds are controlled by multiple women, multiple men, or both a man and a woman.\footnote{Of jointly managed income, in 2010, 1 percent is managed by individuals of the same gender, 1 percent a primary female manager and a secondary male manager, and the remaining 98 percent has a primary male manager and a secondary female manager. In 2013, 2 percent is managed by individuals of the same gender, 3 percent has a primary female manager and a secondary male manager, and the remaining 95 percent has a primary male manager and a secondary female manager. Excluding the 2-5 percent of households which are not male primary, female secondary managed has no significant impact on results. Thus, in the analysis, all shared income is designated as ``joint''.} If no secondary manager is identified, income is classified as either male or female, following the gender of the primary manager. In these definitions, male income is earned by men and female income is earned by women. Joint income is earned by a combination of two individuals, of any gender. 

Considering the differences between these definitions: the primary discrepancy, when joint income is omitted, there is an over-attribution of earned income to men. Women's income remains similar, regardless of whether joint income is included. The percent difference between specifications is approximately 5 percent (on average, 2,828 MK without joint income, compared with 2,686 MK with joint income). In contrast, men's income is vastly different when joint income is included, with a percent difference of around 70 percent (on average, 19,555 MK without joint income, compared with 5,846 MK with joint income). 

Next, I use expenditure data, which also come from the LSMS surveys. An aggregate measure includes all household expenditure, but also include seven measures of disaggregated expenditure for different types of goods. Unsurprisingly, food comprises the largest share of the household's budget on average, followed by housing and utilities expenses.\footnote{This measure generally does not include rent. Of the households included in the analysis, approximately 2 percent do not own their properties, and thus may pay some rent on the building in which they live. Exclusion of renter households does not significantly influence results.} Recreation, hotels and restaurants, and healthcare comprise the smallest budget shares. Table~\ref{tab:sumstat} reports these figures in Panel 2. 

Finally, I use exogenous measures of rainfall to measure the short term shocks which are experienced by households. Table~\ref{tab:sumstat} reports relevant summary statistics in Panel 3. Rainfall data are derived from the CHIRPS rainfall data set, which is produced by the Climate Hazards Group at University of California, Santa Barbara \citep{Funketal15}. CHIRPS makes use of a monthly climatology CHPclim, Tropical Rainfall Measuring Mission Multi-satellite Precipitation Analysis version 7 (TMPA 3B42 v7) and global Thermal Infrared Cold Cloud Duration (TIR CCD) from two NOAA archives. The remote sensing data are then merged with gauge data from five public archives, including the Global Historical Climatology Network (GHCN) and GTS, several private sources, and meteorological agencies. CHRIPS data is measured daily at a 0.05 degree resolution. 

Using data from the LSMS, rainfall can be measured at the household level \citep{MichlerEtAl21}. The data provide a record of rainfall variation across households within a village, although fluctuations within a village for a point in time are relatively small. I use total seasonal rainfall (measured in mm) as my variable of interest. Per \cite{MichlerEtAl21}, I also test a series of other rainfall specifications with no significant differences.  

As rainfall is defined at the household-level, all household members are subject to the same rainfall patterns, but the same level and pattern precipitation may have different effects on the income produced by different crops. That is, a particular rainfall pattern will likely differently impact crops that tend to be produced by women and crops that tend to be produced by men (see Table~\ref{tab:e1cropsmf}). These different impacts are an integral part of our identification strategy, discussed in the subsequent sections.

\section{Theoretical Model}\label{sec:e1theory}

Many household models assume that households reach Pareto efficient outcomes. This assumption is often unchallenged. \cite{Udry96} was among the first to question the fundamental assumption that households in developing countries must be Pareto efficient. Though Udry argues for the rationality of Pareto efficiency, he posits that it is not mandatory. He demonstrates empirically that for farming households in Burkina Faso, Pareto efficient allocation of resources is not achieved. Subsequent research has upheld Udry's findings, demonstrating that many households do not pool income and are not Pareto efficient. Similar behavior is observed in literature on gender relations and bargaining \citep{Duflo03, Udryetal95, Basu06, LiandWu2011, Doss13, Antman15, FialaandHe16}.  

To take these ideas a step further and incorporate ideas of joint and collaborative work, I adapt a model developed by \cite{DufloandUdry04} and include jointly earned income. A household consists of two individuals, each of whom produces one crop on a plot and who together produce a joint crop on a shared plot ($i \in \{m, f, j\}$). Farms are cultivated using labor ($L_i$) which for men ($m$) and women ($f$), can be traded in a competitive market at wage $w$. The agricultural production function is $f_i (L_i, v)$ where $v \equiv (v_1, v_2)^{'}$ is a vector of two measures of rainfall which impact cultivation on the plot of each individual.

Rainfall is experienced and each individual consumes a vector of private goods $x_m$, $x_f$ and some shared public goods $x_p$ at prices $p$. Preferences of men and women follow the expected utility function $\mathrm{E}u_i(x_i, x_p)$, where expectations are taken over potential realizations of rainfall. Rainfall influences the efficient allocation of resources only through its impact on cultivation.\footnote{This is a strong assumption also taken by \cite{DufloandUdry04}, which I explore further in Appendix~\ref{sec:e1pricesec}. If this is the case, non-persistent income variations will not result in changes to the allocation of resources within the household.}

Any \emph{ex ante} efficient allocation of resources can be characterized as a solution to:

\begin{equation}
\begin{aligned}
& \max_{x_i, L_i}
& & \mathrm{E} \big\{ [u_f(x_f, x_p)]  + \Lambda [u_m(x_m, x_p)]  \big\} \\ 
& \text{s.t.} 
& & p \cdot (x_m + x_f + x_p) \le f_f(L_f, v) + f_m(L_m, v) + f_j(L_j, v) - w(L_f + L_m), 
\end{aligned}
\end{equation}

\noindent where $\Lambda$ represents some Pareto weight or bargaining power, which depends on the observable and unobservable attributes of household members. This Pareto weight does not depend on $v$, as with an efficient allocation of resources, risk is pooled. Recall that, if risk is pooled, households insure one another against rainfall variation and over the period of interest and thus there is no a change in bargaining power.

With expenditure as $e \equiv p \cdot (x_m + x_f + x_p)$: 

\begin{equation}\label{eq:e1denotex}
x_i = x_i( \Lambda, p, e) \ \ \ \ \ \ \forall \ \  i \in \{m, f\}. 
\end{equation}

\noindent Consumption of any particular good is independent of the rainfall realization $v$, conditional on expenditures, prices, preferences, and the Pareto weight parameter. So, the impact of rainfall realizations on expenditure for a particular commodity depends only on the expenditure elasticity of demand for that commodity and on the effect of rainfall on overall expenditure.

Equation~\eqref{eq:e1denotex} implies that the impact of rainfall realizations on expenditure for any particular commodity depends only on the expenditure elasticity of demand for that commodity and on the effect of rainfall on overall expenditure. And so, assuming that the relative prices of consumption are not related to rainfall realizations ($\frac{\partial p}{\partial v} = 0$).

For any $i$ and period $t \in \{1, 2\}$ and any good $g$:

\begin{equation}\label{eq:e1restrict1}
\frac{dx_{i}^{g}}{dv_t} = \frac{dx_{i}^{g}}{de} \cdot \frac{de}{dv_t}. 
\end{equation}

\noindent Equation~\eqref{eq:e1restrict1} demonstrates the effect of rainfall in period $t$ on consumption of good $g$ by $i$. Its impact on total expenditure should be equal across all rainfall realizations: 

\begin{equation}\label{eq:e1doublerainratio}
\frac{\frac{dx_{i}^{g}}{dv_1}}{\frac{de}{dv_1}} = \frac{\frac{dx_{i}^{g}}{dv_2}}{\frac{de}{dv_2}}. 
\end{equation}

\noindent What is crucial in equation~\eqref{eq:e1doublerainratio} is that $dv_t$ impacts collective household decision making through its influence on the household's budget constraint.\footnote{Only data on rainfall and expenditures is required in order to estimate equations~\eqref{eq:e1denotex} and \eqref{eq:e1doublerainratio}. $f_f(L_f, r) + f_m(L_m, r) + f_j(L_j, r) - w(L_f + L_m)$ is not observed and such data is not required for empirical analysis.} 

Equation~\eqref{eq:e1doublerainratio} serves as an overidentifying restriction, which is tested in the empirical analysis. The restriction specifies that realized rainfall influences demand for a particular good in proportion to its impact on expenditure. 

\section{Empirical Implementation and Identification}\label{sec:e1subsecimplement}

To implement the overidentifying restriction test in equation~\eqref{eq:e1doublerainratio}, a log-linear approximation of demand is taken for a certain commodity $g$ by household $h$ in period $t$: 

\begin{equation}\label{eq:e1implement1}
\log(x_{ght}) = \alpha \cdot \log(e_{ht}) + f(\Lambda) + Z_{ht} \beta + \gamma_h + \nu_{ht}
\end{equation}

\noindent where $e_{ht}$ again denotes expenditure, while $Z_{ht}$ represents location and time indicators, $\gamma_h$ represents a household fixed effect, and $\nu_{ht}$ represents an error term. Markets are assumed to be regionally integrated. With this assumption, at any point in time, prices are the same within a region and so is the effect of rainfall on relative prices. Demand is captured by the region-time indicator. 

Next assuming a log-linear relationship between total expenditure and profit, for the rainfall vector, the relationship between rainfall and household expenditure (total and on a particular commodity) is represented: 

\begin{equation}\label{eq:e1implement2}
\log(e_{ht}) = V_{ht}\alpha + Z_{ht} \beta + \epsilon_{ht} 
\end{equation}

From this, I can specify the reduced form relationship: 

\begin{equation}\label{eq:e1implement3}
	\log(x_{hgt}) = V_{ht} \pi + f(\Lambda) + X_{ht} \beta + \gamma_h + \nu_{ht} 
\end{equation}

\noindent Equation~\eqref{eq:e1implement2} and \eqref{eq:e1implement3} are differenced, giving reduced form equations for estimation:

\begin{equation}
	\Delta \log(e_{hit}) = \Delta V_{ht} \alpha + \Delta X_{ht} \beta + \Delta \epsilon_{ht}
\end{equation}

\begin{equation}
	\Delta \log(x_{hgt}) = \Delta V_{ht} \pi + \Delta X_{ht} \beta + \Delta \nu_{ht}  
\end{equation}

Equation~\eqref{eq:e1implement2} allows for the analysis of changes over time, and controls for unobserved household heterogeneity, which, if neglected, could bias coefficient estimates. To test the restriction proposed in equation~\eqref{eq:e1doublerainratio}, I use the following overidentification test: 

\begin{equation}
	f_f(L_f, v) + f_m(L_m, v) + f_j(L_j, v) - w(L_f + L_m) = \chi \alpha
\end{equation}

\noindent for some scalar $\chi$. To test this empirically, I employ a proportional non-linear Wald test. However, this test is limited as it does not explicitly link variation in income with its respective gendered or joint origin.\footnote{Results of the unrestricted test are presented in Table~\ref{tab:e1unconestimates} in Appendix~\ref{sec:unrestricoverid}.} 

To address this limitation, I generate linear differences in rainfall for each of the income earners within a household. First differenced income ($y_{iht}$) regressions are estimated for the first stage:

\begin{equation}\label{eq:e1forpredvalues}
\Delta log(y_{iht}) =  \Delta V_{ht} \psi_{yi} +  \Delta Z_{ht} \delta_{yi} +  \Delta \gamma_{it}. 
\end{equation}

\noindent And calculate predicted values ($ \Delta V_{ht} \hat{\psi}_{yi}$). Then, I estimate for the second stage:  

\begin{equation}\label{eq:e1expdiff}
\Delta \log(e_{ht}) = \sum_{g=1}^{G} \Delta V_{ght} \hat{\psi}_{gyh} + \Delta Z_{ht} \beta + \Delta \epsilon_{ht},
\end{equation}

\noindent is estimated, where terms are as previously defined. Equation \eqref{eq:e1expdiff} allows for the testing of the impact of rainfall on expenditure, distinguished by different income sources. Instead of broadly considering rainfall's impact on expenditure. I estimate its impact while also testing assumptions of household behavior. 

In implementing this analysis, I control for heteroskedasticity and correlation within households using a clustered bootstrap procedure at the household level. 

\section{Results and Discussion}

In this section, I present and discuss the results of the traditional and extended specifications, as well as several robustness checks on those findings. First, I discuss the results and compare the extended and traditional specifications which include and do not include joint income earned by multiple earners, respectively, for my standard overidentification test. These results are presented in Tables~\ref{tab:e1firststagemfmfj} and \ref{tab:e1overidmfmfj}. I then also discuss two additional robustness checks, which account for heterogeneity across household type and cultures. These results are presented in Tables~\ref{tab:e1s2estimatesmatri} and \ref{tab:e1s2estimatesfem}. 

\subsection{Extended and Traditional Specifications}

I first consider the results of the first stage, the relationship between income and rainfall, presented in Table~\ref{tab:e1firststagemfmfj}. Panel 1 presents the specification with inclusion of male, female, and joint income, while Panel 2 presents the specification with just male and female income.\footnote{Appendix section~\ref{sec:unrestricoverid} presents and discusses the results if this stage did not exist.} In both panels, it is evident that total seasonal rainfall has a positive and significant, albeit small, impact on incomes. In Panel 1, all income types all have a positive relationship with total seasonal rainfall, although the coefficient on male income is more than twice as large as the coefficients on joint and female income. This suggests - following expectations - that as rainfall increases, incomes increase and as rainfall decreases, incomes decrease. 

Interestingly, in Panel 2, the coefficient on female income is more than twice as large as the coefficient on male income. If we assume that larger coefficients are an indicator of greater response to changes in total rainfall, then this suggests that when joint income is not included, as in the traditional specification, men's income becomes less vulnerable to, but also less able to benefit from, fluctuations and short term shocks in seasonal rainfall. Similarly, this suggests that women's income becomes more vulnerable to and more able to benefit from the same fluctuations. 

Using the predicted values from the first stage, I estimate the restricted overidentification test that is the second stage. The results are presented in Table~\ref{tab:e1overidmfmfj}. Recall that the impacts on income by earner, measured conditional on total expenditure, determine discrepancies and shifts in the allocation of income to different expenditure purposes within the household. Panel 1 presents the results of the extended specification which includes joint income. Considering these results, I fail to reject equality for the case of food. Failure to reject equality indicates that income is pooled for this expenditure and that households insure one another for expenditures on goods in this category. That is, I find evidence that households are insuring one another against short term shocks for expenses that the household makes on food. As this comprises a large portion of the household budget and percentage of total household expenditures, this represents a significant expense for the household that is insured. Still, though, I reject equality for expenditures on alcohol and tobacco, clothing, recreation, education, healthcare, housing and utilities, transportation, communication, and hotels and restaurants. That is, all other household expenses. This suggests a strategic - though incomplete - pooling mechanism employed by households. 

I also consider the size and significance of coefficients presented in Panel 1, with particular attention to the magnitudes of these coefficients. The coefficients are interpreted as percentages changes in expenditure, with related changes in income. I observe that there are more significant coefficients associated with predicted changes in women's income, then with men's income or with joint income. About 80 percent of coefficients are significant, suggesting that expenditures on various goods are vulnerable to the predicted changes in women's income. These coefficients are all positive and indicate shifts between 2.5 and 5 percent, depending on the expenditure. This suggests potentially consequential changes in expenditure to various good categories, as women's income changes, though not necessarily for men's income or income from joint sources. 

In Panel 2, the results of the traditional specification are presented, which does not include joint income. Similarly to the extended specification, I fail to reject equality for the case of food. Again: failure to reject equality indicates that income is pooled for this expenditure and households insure one another for expenditures on goods in this category. Further, as before, I reject equality for expenditures on alcohol and tobacco, clothing, recreation, education, healthcare, housing and utilities, transportation, communication, and hotels and restaurants. Overall, this indicates that the findings do not differ across the extended and traditional specification, at least with respect to the broad conclusions about insurance of expenditures by household members in the presence of short term shocks.

It is worth noting, however, that while the overall interpretation of results are the same, there are differences in the magnitude and significance of coefficients across the two specifications. In the traditional specification, only 40 percent of the predicted changes in women's income have a significant relationship with expenditure - half as many as in the extended specification. Further, all of the magnitudes are greater in the traditional specification. Considering the implications of these differences, this change suggests that actual household behaviors may be masked to some extent in the traditional specification. While to some degree, the predicted changes, particularly in women's income, suggest a smaller percentage change in magnitude of the coefficients when joint income is included, but a greater number of categories of expenditure overall show a significant change in predicted income and corresponding changes in expenditure. This suggests that there is a shift in type of allocation behavior demonstrated, that is important for understanding dynamics and functioning of households and their responses to coping with short term shocks and fluctuations in income. It suggests that women's income changes result in changes to expenditure - if women's incomes increase, expenditure increases, but if women's income decreases, expenditure decreases. This is observed particularly for healthcare and transportation, which see no changes in expenditure associated with changes in income from men or joint sources. The differences observed in allocation behavior between Panel 1 and 2 suggests that the complete story of household allocation behavior is not evident in the traditional specification, but is only revealed in the extended specification, which includes income earned jointly.

These results suggest the importance in and value for including joint relationships in analyses of household behavior and resource allocation. The evidence of this study elucidates and differs from previous work, and moves beyond the usual conceptions of the household, extending the discussion about intra-household relationships and behavior. Although I find evidence of partial income pooling, that behavior is likely strategic, as it is for the greatest single expenditure source within most households' budgets. In addition to this, however, the findings in this section suggest that understanding detailed intra-household interactions and their outcomes on household behavior have important consequences for household resource allocation and decision making, as related to income allocation in the presence of shocks. Understanding the dynamics of household members' allocation of resources to various expenditures is essential for appreciating how decisions are made within households and how members individually and households at large, cope with shocks to their incomes. 

\subsection{Heterogeneity across Households}\label{sec:e1types}

Next, I present two robustness checks. To consider household composition may change the allocation of household resources, I consider some potential alternative household structures, including households with matrilineal traditions and households with female headship. It is possible that a subset of households may behave in different ways based on different compositions and structures, which are consistent with disparate types of income pooling behaviors. 

In Table~\ref{tab:e1s2estimatesmatri}, Panel 1 presents the male, female, and joint specification, while Panel 2 presents simply the male and female specification. Within each panel, there are two sections. In each, section A presents the results for non-matrilineal households while section B presents the results for matrilineal households. This is similar in Table~\ref{tab:e1s2estimatesfem}, though in this case, section A presents the results for non-female-headed households and section B presents the results for female-headed households. 

First, I consider differences in income pooling driven by societal characteristics, in particular matrilineal inheritance traditions. \cite{Walther16} finds that there may be disparities in cooperative decision making behavior in Malawi, depending on women's status in the household. Specifically, those residing in communities with matrilineal lineage may exhibit different behavior, due to women's relative power in these communities. To explore this, I consider whether the community reports being matrilineal. The data includes responses to the question: ``Do individuals in this community trace their descent through their father, their mother, or are both kinds of descent traced?'' Communities which respond ``their mother'' are deemed to be matrilineal.\footnote{Less than 10 percent of respondents indicated ``both'' and so these communities were grouped with those communities which trace lineage from their father. This entire group is simply classified as ``non-matrilineal''.} 

The results of the specifications related to matrilineal inheritance traditions are presented in Table~\ref{tab:e1s2estimatesmatri}. In the table, Panel 1 presents the extended specification which includes male, female, and joint incomes, while Panel 2 presents the traditional specification, which includes only male and female incomes. Within each panel, there are two sections. In each, section A presents the results for non-matrilineal households while section B presents the results for matrilineal households. As in both the extended and traditional specification presented in Table~\ref{tab:e1overidmfmfj}, I fail to reject equality for the case of food. Failure to reject equality indicates that income is pooled for this expenditure and that households insure one another for expenditures on goods in this category. This is the case for all variations, including matrilineal and non-matrilineal households, in both specifications. Again, I reject equality for expenditures on alcohol and tobacco, clothing, recreation, education, healthcare, housing and utilities, transportation, communication, and hotels and restaurants. This may a strategic pooling behavior employed by households to pool income for their greatest household expenditure, although the overall household expenditure pooling is thus incomplete. 

I observe some differences when examining matrilineal households in the different specifications. In matrilineal households in the extended specification, predicted changes in women's income are significantly associated with expenditures in about half of the cases; but this falls to just about 40 percent in the non-matrilineal specification. However, in matrilineal households, very few joint income coefficients are significant, while in the non-matrilineal households, more than 30 percent are significant. Notably, coefficients are also smaller in the non-matrilineal specification, between 2 and 4 percent, rather than 3 to 6 percent as in the matrilineal specification. Taking this further, in the traditional specification, in non-matrilineal households, two thirds of the coefficients are significant for women's income, while only 25 percent are significant for men's income. Again, this falls in the matrilineal specification: just over 15 percent are significant for women's income, while none are significant for men's income. This seems to suggest that there are disparities in results, with respect to the magnitude and significance of coefficients, when considering the traditional and extended specifications, and, as such, conclusions may ultimately vary based on what specification is considered. In particular, some specifications, particularly when considering matrilineal households, may hide women's allocation of resources in response to short term shocks.

The results of the specification related to male and female household headship are presented in Table~\ref{tab:e1s2estimatesfem}. In the table, Panel 1 presents the extended specification which includes male, female, and joint incomes, while Panel 2 presents the traditional specification, which includes only male and female incomes. Within each panel, there are two sections. In each, section A presents the results for non-female-headed households while section B presents the results for female-headed households. Once again, I fail to reject equality for the case of food. Failure to reject equality indicates that income is pooled for this expenditures and that households insure one another for expenditures on goods in this category. This is the case for all variations, including matrilineal and non-matrilineal households, in both specifications. And again, I reject equality for expenditures on alcohol and tobacco, clothing, recreation, education, healthcare, housing and utilities, transportation, communication, and hotels and restaurants. There is one exception to this finding: in the case of female-headed households in the extended specification. In this case, we fail to reject equality for the case of hotels and restaurants. This relationship however, is likely spurious, simply based on the consistency of results throughout the rest of the analysis. 

In the case of female- and male-headed households, I see very similar results to the findings for matrilineal and non-matrilineal households: predicted changes in women's income are significantly associated with expenditures in two thirds of the non-female-headed specification, this falls to just over 15 percent in the female-headed specification. Interestingly, in neither case for the extended specification are the joint incomes subject to many significant relationships. Coefficient are also very large in the non-female-headed specification, with coefficients between 9 and 17 percent, relative to the female-headed specification, with coefficients around 1 or 2 percent. This suggests that in non-female-headed households, changes in women's income due to positive or negative short term shocks are associated with changes to expenditure - positive or negative, depending on the direction of the shock. Further, in the traditional specification, in both the female and non-female-headed household specifications only about 15 percent of coefficients for women's income are significant. And, it is not the case that coefficients on male incomes are significant instead. In both cases, very little is significant. Again, this suggests that the traditional specification obfuscates and masks some information about how households cope within households and how expenditures are adjusted, after the experience of a short term shock. The specification used in an analysis may obfuscate the role of women in household resource allocation and understanding household expenditures, in response to shock events. 

It is worth noting that the differences in these robustness checks, differ somewhat from previous literature. Previous work suggests that there are likely to be significant differences based on lineage and household headship. But, I find little evidence of that, with respect to intra-household resource allocation. I suspect that the  context of Malawi may be important to consider: more than two thirds of households report a matrilineal heritage, but still nearly 80 percent of households are male-headed. This configuration may explain some of the disparities from previous literature, at least with respect to these robustness checks focused on household structures and traditions. 

The differences in potential expenditure changes revealed by considering both the extended and traditional specifications demonstrates the vulnerabilities of and opportunities to benefit in households' livelihoods associated with changes in women's incomes, resulting from these short term shocks. Shocks and resulting changes to women's income is significantly affects expenditure more often then changes in men's and joint income. This is a fact often hidden in the traditional specification when only men's and women's incomes are included. These changes are likely to affect household expenditures and, in turn, impact household livelihoods and outcomes as a result. The results reveal a partial, strategic pooling of household incomes on food - the largest expenditure within most households' budgets. However, overall, households do not pool income completely. These findings help further to elucidate intra-household behavior. Understanding these dynamics and the full picture of household vulnerabilities is revealed only in the extended specification, but is imperative for fully understanding household dynamics and intra-household resource allocation in the presence of short term, transient shocks.

\section{Conclusions}

I reject the hypothesis of complete income pooling within the household when income earned jointly by two or more individuals is included in the empirical analysis, I find evidence that household members partially insure one another for expenditures on food. Food is a significant component of a household's expenditure and so this pooling behavior is for large percentage of a household's shared expenditures. This may a strategic pooling behavior employed by households to pool income for their greatest household expenditure, although the overall household pooling is still incomplete.  But, this finding is important as they are generally contrary to previous studies, which fail to find even partial pooling or insurance within households. This finding is robust across all of the specifications considered in this paper, including the traditional and extended specifications, as well as robustness checks of household structure, considering dynamics of matrilineal households and households headed by women. 

The inclusion of joint income reveals vulnerabilities of  and opportunities for households not previously revealed in most traditional analyses of household behavior. Short term, non-persistent shocks often result in a predicted change in women's income, which is positively associated with changes in various household expenditures. This suggests that as women's income increases, household expenditures also increase. However, it further suggests that if women's income decline, then expenditures may decrease - and these expenditures are not insured by other household members and so may result in shifts in the overall household expenditure. This is an important dynamic of household resource allocation, which is not evident in a binary, traditional conception of the collective household model, but is evident when joint income is fully included in the analysis.   

Household models have been a cornerstone of microeconomic analysis for more than half a century, but many of the analyses on intra-household allocation of resources has assumed that all income is earned by a single household member, who is then responsible for those earnings in intra-household resource allocation. In these circumstances, findings indicate that households tend not to pool income. In this paper, I have relaxed this assumption and reveal differential household behavior, not previously shown in standard, traditional models, which only consider men and women and their income in the analysis. In this way, this study moves beyond standard unitary, collective, and non-cooperative conceptions of the household, extending the existing discussion about intra-household dynamics. In contrast with previous studies and prevailing empirical practice, the results suggest that understanding joint interactions and their outcomes on household behavior have important consequences for households and are imperative for our understanding of households dynamics and interactions. Consideration of these relationships, combined with efforts to learn about the intricate, shared relationships within households will be imperative for understanding those dynamics. 

\newpage
\singlespace
\bibliographystyle{chicago} 
\bibliography{uptodatedissbib}

\newpage 

\begin{table}[!htbp]\centering
	\def\sym#1{\ifmmode^{#1}\else\(^{#1}\)\fi}
	\caption{Crop Cultivation, By Management Type \label{tab:e1cropsmf}}
	\resizebox{5in}{!}
	{\setlength{\linewidth}{.1cm}\newcommand{\contents}
		{\begin{tabular}{l*{7}{D{.}{.}{-1}}}
				\\[-1.8ex]\hline 
				\hline \\[-1.8ex]
				&\multicolumn{3}{c}{2010} &\multicolumn{3}{c}{2013}  \\
				\midrule 
				&\multicolumn{1}{c}{	\hspace{0.1cm} Women} &\multicolumn{1}{c}{ 	\hspace{0.1cm}Men} &\multicolumn{1}{c}{	\hspace{0.1cm} Joint} &\multicolumn{1}{c}{ 	\hspace{0.1cm} Women} &\multicolumn{1}{c}{ 	\hspace{0.1cm} Men} &\multicolumn{1}{c}{ 	\hspace{0.1cm} Joint}  \\
				\midrule
				\hspace{0.1cm} Maize  & 30.93 & 28.61 & 29.41  & 25.20 & 25.58 & 25.39 \\ 
				\hspace{0.1cm} Tobacco  & 7.18 & 25.85 & 24.37 & 4.46 & 12.00 & 11.99 \\ 
				\hspace{0.1cm} Groundnut & 21.0 & 17.69 & 19.32 & 23.17 & 22.79 & 23.82 \\ 
				\hspace{0.1cm} Rice & 2.75 & 2.0 & 2.24 & 3.26 & 2.41 & 2.85   \\ 
				\hspace{0.1cm} Beans & 20.99 & 14.45 & 13.16 & 26.84 & 20.37 & 20.35 \\ 
				\hspace{0.1cm} Potato & 3.31 & 3.07 & 3.36 & 3.26 & 3.15 & 3.00 \\ 
				\hspace{0.1cm} Millet  & 1.66 & 0.62 & 0.84 & 0.00 & 0.24 & 0.16 \\ 
				\hspace{0.1cm} Sorghum  & 1.10 & 0.31 & 0.56 & 2.03 & 0.97 & 0.79 \\ 
				\hspace{0.1cm} Other & 11.08 & 7.4 & 6.29 & 11.78 & 12.49 & 11.65 \\ 
				\hline  
				\hline \\[-1.8ex] 
				\multicolumn{7}{p{\linewidth}}{\footnotesize \textit{Note}: Table reports the crops cultivated, classified by management type. Management type includes ``male'', ``female'', or ``joint'', following the joint specification as describe in section~\ref{sec:data}.   Reallocated and omitted specifications are not included.  Maize includes: local maize and hybrid maize. Beans include: ``beans'', soybeans, and pigeonpea. Other crops include: cotton, sugarcane, cotton, tomato, okra, and various vegetables and spices.} \\
		\end{tabular}}
		\setbox0=\hbox{\contents}
		\setlength{\linewidth}{\wd0-2\tabcolsep-.25em}
		\contents}
\end{table}	

\newpage

\begin{table}[!htbp]\centering
	\caption{Summary Statistics for Income, Expenditure, and Rainfall \label{tab:sumstat}}
	\resizebox{5.5in}{!}
	{\begin{tabular}{l*{3}{D{.}{.}{-1}}}
			\\[-1.8ex]\hline 
			\hline \\[-1.8ex] 
			&\multicolumn{1}{c}{Mean} &\multicolumn{1}{c}{Std Dev} \\
			\midrule
			\multicolumn{3}{c}{\textbf{\emph{Panel 1: Income}}} \\ 
			\midrule 
			\hspace{0.1cm} \textbf{Joint Income} && \\
			\hspace{0.1cm} Female Income  & 2,686 & 53,023  \\
			\hspace{0.1cm} Male Income  & 5,846 & 39,134  \\ 
			\hspace{0.1cm} Joint Income  & 13,871 & 103,771   \\ 
			\hspace{0.1cm} \textbf{Omitted Joint Income} && \\
			\hspace{0.1cm} Female Income & 2,828  & 53,325  \\ 
			\hspace{0.1cm} Male Income & 19,555 & 111,606  \\ 
			\hline
			\midrule  
			\multicolumn{3}{c}{\textbf{\emph{Panel 2: Expenditures}}} \\ 
			\midrule
			\hspace{0.1cm} Total & 611,834 & 670,837 \\
			\hspace{0.1cm} Food  & 359,981 & 318,225 \\
			\hspace{0.1cm} Cigarettes and Alcohol  & 17,746 & 47,288 \\
			\hspace{0.1cm} Clothing & 20,548 & 47,523 \\
			\hspace{0.1cm} Recreation & 4,664 & 17,225 \\ 
			\hspace{0.1cm} Education & 9,847 & 41,409 \\
			\hspace{0.1cm} Healthcare  & 7,709 & 18,966 \\
			\hspace{0.1cm} Housing and Utilities & 119,296 & 167,730 \\
			\hspace{0.1cm} Transportation & 30,182 & 107,582 \\
			\hspace{0.1cm} Communication & 19,722 & 94,566 \\
			\hspace{0.1cm} Hotels and Restaurants & 7,012  & 17,200  \\ 
			\hline
			\midrule  
			\multicolumn{3}{c}{\textbf{\emph{Panel 3: Rainfall}}} \\ 
			\midrule 
			\hspace{0.1cm} Total Seasonal Rainfall & 1,163 & 551 \\ 
			\hline \\[-1.8ex] 
			\multicolumn{2}{c}{\footnotesize \textit{Note}: Income and expenditure values are in Malawian Kwacha real terms with 2010 base. Rainfall measured in millimeters.} \\
	\end{tabular}}	
\end{table}  

\newpage

\begin{table}[htbp]\centering
	\def\sym#1{\ifmmode^{#1}\else\(^{#1}\)\fi}
	\caption{First Stage Rainfall Estimates \label{tab:e1firststagemfmfj}}
	\resizebox{4.5in}{!}
	{\setlength{\linewidth}{.1cm}\newcommand{\contents}
		{\begin{tabular}{l*{4}{D{.}{.}{-1}}}
				\\[-1.8ex]\hline 
				\hline \\[-1.8ex]
				&\multicolumn{1}{c}{\hspace{0.1cm} Joint Income} &\multicolumn{1}{c}{\hspace{0.1cm} Female Income} &\multicolumn{1}{c}{\hspace{0.1cm} Male Income}\\
				\hline 
				\midrule
					\multicolumn{4}{c}{\textbf{\emph{Panel 1: Male, Female, and Joint}}} \\ 
					\midrule 
				\hspace{0.1cm} Total Seasonal Rainfall    &     0.00039\sym{*} &     0.00045\sym{**}&     0.00113\sym{**}\\
				&   (0.00016)        &   (0.00014)        &   (0.00022)        \\
				\midrule
				\hspace{0.1cm} \(R^{2}\)        &\multicolumn{1}{c}{0.020}        &\multicolumn{1}{c}{0.022}        &\multicolumn{1}{c}{0.048}        \\
	
					\hline 
				\midrule
				\multicolumn{4}{c}{\textbf{\emph{Panel 2: Male and Female}}} \\ 
				\midrule 
				\hspace{0.1cm} Total Seasonal Rainfall &   &     0.00167\sym{**}&     0.00050\sym{**}\\
			&	&   (0.00024)        &   (0.00014)        \\
				\midrule
				
					\hspace{0.1cm} \(R^{2}\)       &   &\multicolumn{1}{c}{0.075}        &\multicolumn{1}{c}{0.023}        \\
				\hline 
				\hline \\[-1.8ex] 
				\multicolumn{4}{p{\linewidth}}{\footnotesize \textit{Note}: N = 850. Fully robust standard errors are in parentheses. (\sym{*} \(p<0.05\), \sym{**} \(p<0.01\)). Regressions also include agro-ecological zone indicators.} \\
		\end{tabular}}
		\setbox0=\hbox{\contents}
		\setlength{\linewidth}{\wd0-2\tabcolsep-.25em}
		\contents}
\end{table}

\begin{landscape}	
	\begin{table}[htbp]\centering
		\def\sym#1{\ifmmode^{#1}\else\(^{#1}\)\fi}
		\caption{Restricted Overidentification Tests \label{tab:e1overidmfmfj}}
		\resizebox{9in}{!}
		{\setlength{\linewidth}{.1cm}\newcommand{\contents}
			{\begin{tabular}{l*{12}{D{.}{.}{-1}}}
					\\[-1.8ex]\hline 
					\hline \\[-1.8ex]
			   &\multicolumn{1}{c}{(1)}&\multicolumn{1}{c}{(2)}&\multicolumn{1}{c}{(3)}&\multicolumn{1}{c}{(4)}&\multicolumn{1}{c}{(5)}&\multicolumn{1}{c}{(6)}&\multicolumn{1}{c}{(7)}&\multicolumn{1}{c}{(8)}&\multicolumn{1}{c}{(9)}&\multicolumn{1}{c}{(10)}&\multicolumn{1}{c}{(11)}\\
				&\multicolumn{1}{c}{Aggregate}&\multicolumn{1}{c}{Food}&\multicolumn{1}{c}{Alcohol \& Tobacco}&\multicolumn{1}{c}{Clothing}&\multicolumn{1}{c}{Recreation}&\multicolumn{1}{c}{Education}&\multicolumn{1}{c}{Healthcare}&\multicolumn{1}{c}{Housing \& Utilities }&\multicolumn{1}{c}{Transportation}&\multicolumn{1}{c}{Communication}&\multicolumn{1}{c}{Hotels \& Restaurants}\\
					\hline 
				\midrule
				\multicolumn{12}{c}{\textbf{\emph{Panel 1: Male, Female, and Joint}}} \\ 
				\midrule 
				\hspace{0.1cm} Predicted change in male income&       5.008\sym{**}&       5.029\sym{*} &      -0.142        &      -2.603        &       3.568\sym{*} &       3.113\sym{*} &       1.043        &       3.944\sym{*} &       1.160        &       0.145        &       2.180        \\
				&     (1.921)        &     (2.021)        &     (1.814)        &     (2.026)        &     (1.462)        &     (1.534)        &     (1.918)        &     (1.555)        &     (1.891)        &     (1.679)        &     (1.729)        \\
				\hspace{0.1cm} Predicted change in female income&       4.127\sym{**}&       4.176\sym{**}&       1.664        &       2.666\sym{**}&       3.006\sym{**}&       2.985\sym{**}&       2.585\sym{**}&       3.007\sym{**}&       2.170\sym{*} &       2.112\sym{*} &       0.455        \\
				&     (0.953)        &     (1.003)        &     (0.900)        &     (1.005)        &     (0.725)        &     (0.761)        &     (0.951)        &     (0.771)        &     (0.938)        &     (0.833)        &     (0.858)        \\
				\hspace{0.1cm} Predicted change in joint income&       0.601        &       0.570        &       0.865        &       2.045\sym{*} &      -1.054        &      -0.574        &       0.354        &       0.224        &       0.586        &       0.239        &       0.237        \\
				&     (0.848)        &     (0.892)        &     (0.801)        &     (0.895)        &     (0.645)        &     (0.678)        &     (0.847)        &     (0.687)        &     (0.835)        &     (0.742)        &     (0.763)        \\
				\midrule
					
				\hspace{0.1cm}  Overidentification - Wald Test
				&\multicolumn{1}{c}{}         
				&\multicolumn{1}{c}{0.26}      
				&\multicolumn{1}{c}{73.25}      
				&\multicolumn{1}{c}{215.62}         
				&\multicolumn{1}{c}{206.78}         
				&\multicolumn{1}{c}{109.73}         
				&\multicolumn{1}{c}{35.58}         
				&\multicolumn{1}{c}{24.38}         
				&\multicolumn{1}{c}{40.63}         
				&\multicolumn{1}{c}{69.68}         
				&\multicolumn{1}{c}{63.82}        \\
				&\multicolumn{1}{c}{}         
				&\multicolumn{1}{c}{(0.968)}      
				&\multicolumn{1}{c}{(0.000)}      
				&\multicolumn{1}{c}{(0.000)}         
				&\multicolumn{1}{c}{(0.000)}         
				&\multicolumn{1}{c}{(0.000)}         
				&\multicolumn{1}{c}{(0.000)}         
				&\multicolumn{1}{c}{(0.000)}         
				&\multicolumn{1}{c}{(0.000)}         
				&\multicolumn{1}{c}{0(.000)}         
				&\multicolumn{1}{c}{(0.000)}        \\
					
					\hspace{0.1cm}  \(R^{2}\)          &\multicolumn{1}{c}{0.214}        &\multicolumn{1}{c}{0.196}        &\multicolumn{1}{c}{0.064}        &\multicolumn{1}{c}{0.080}        &\multicolumn{1}{c}{0.060}        &\multicolumn{1}{c}{0.075}        &\multicolumn{1}{c}{0.061}        &\multicolumn{1}{c}{0.201}        &\multicolumn{1}{c}{0.060}        &\multicolumn{1}{c}{0.031}        &\multicolumn{1}{c}{0.056}        \\
	
		\hline 
	\midrule
	\multicolumn{12}{c}{\textbf{\emph{Panel 2: Male and Female}}} \\ 
	\midrule 
	
\hspace{0.1cm} Predicted change in male income&       0.228        &       0.190        &       0.872        &       2.229\sym{*} &      -1.488        &      -0.935        &       0.204        &      -0.083        &       0.481        &       0.140        &       0.148        \\
&     (1.049)        &     (1.103)        &     (0.990)        &     (1.106)        &     (0.798)        &     (0.838)        &     (1.047)        &     (0.849)        &     (1.032)        &     (0.917)        &     (0.944)        \\
\hspace{0.1cm} Predicted change in female income&       8.229\sym{*} &       8.347\sym{*} &       0.418        &      -2.481        &       8.097\sym{**}&       6.954\sym{*} &       3.258        &       6.578\sym{*} &       2.575        &       2.084        &       2.158        \\
&     (3.576)        &     (3.762)        &     (3.377)        &     (3.771)        &     (2.721)        &     (2.856)        &     (3.570)        &     (2.894)        &     (3.520)        &     (3.126)        &     (3.218)        \\

	\midrule
	\hspace{0.1cm}  Overidentification - Wald Test
	&\multicolumn{1}{c}{}         
	&\multicolumn{1}{c}{0.19}      
	&\multicolumn{1}{c}{72.20}      
	&\multicolumn{1}{c}{211.56}         
	&\multicolumn{1}{c}{205.81}         
	&\multicolumn{1}{c}{109.57}         
	&\multicolumn{1}{c}{34.10}         
	&\multicolumn{1}{c}{24.23}         
	&\multicolumn{1}{c}{39.97}         
	&\multicolumn{1}{c}{67.02}         
	&\multicolumn{1}{c}{63.24}        \\
	&\multicolumn{1}{c}{}         
	&\multicolumn{1}{c}{(0.909)}      
	&\multicolumn{1}{c}{(0.000)}      
	&\multicolumn{1}{c}{(0.000)}         
	&\multicolumn{1}{c}{(0.000)}         
	&\multicolumn{1}{c}{(0.000)}         
	&\multicolumn{1}{c}{(0.000)}         
	&\multicolumn{1}{c}{(0.000)}         
	&\multicolumn{1}{c}{(0.000)}         
	&\multicolumn{1}{c}{(0.000)}         
	&\multicolumn{1}{c}{(0.000)}        \\
	\hspace{0.1cm}  \(R^{2}\)        &\multicolumn{1}{c}{0.224}        &\multicolumn{1}{c}{0.200}        &\multicolumn{1}{c}{0.080}        &\multicolumn{1}{c}{0.076}        &\multicolumn{1}{c}{0.056}        &\multicolumn{1}{c}{0.076}        &\multicolumn{1}{c}{0.053}        &\multicolumn{1}{c}{0.197}        &\multicolumn{1}{c}{0.068}        &\multicolumn{1}{c}{0.028}        &\multicolumn{1}{c}{0.054}        \\
	
					\hline 
					\hline \\[-1.8ex] 
					\multicolumn{12}{p{\linewidth}}{\footnotesize  \textit{Note}: N = 850. The table presents coefficients of the difference in log consumption of each item on the difference in predicted log income, as from Table~\ref{tab:e1firststagemfmfj}. Regressions include agro-ecological zone indicators. Fully robust standard errors clustered at the household are in parentheses  (\sym{*} \(p<0.05\), \sym{**} \(p<0.01\)) below the predicted change variables. Below the Wald test in parentheses is the probability $> \chi^2$.} \\
			\end{tabular}}
			\setbox0=\hbox{\contents}
			\setlength{\linewidth}{\wd0-2\tabcolsep-.25em}
			\contents}
	\end{table}

\end{landscape}

\begin{landscape}	
	\begin{table}[htbp]\centering
		\def\sym#1{\ifmmode^{#1}\else\(^{#1}\)\fi}
		\caption{Restricted Overidentification Tests - Log of Consumption (Matrilineal) \label{tab:e1s2estimatesmatri}}
		\resizebox{9in}{!}
		{\setlength{\linewidth}{.1cm}\newcommand{\contents}
			{\begin{tabular}{l*{12}{D{.}{.}{-1}}}
					\\[-1.8ex]\hline 
					\hline \\[-1.8ex]
				 &\multicolumn{1}{c}{(1)}&\multicolumn{1}{c}{(2)}&\multicolumn{1}{c}{(3)}&\multicolumn{1}{c}{(4)}&\multicolumn{1}{c}{(5)}&\multicolumn{1}{c}{(6)}&\multicolumn{1}{c}{(7)}&\multicolumn{1}{c}{(8)}&\multicolumn{1}{c}{(9)}&\multicolumn{1}{c}{(10)}&\multicolumn{1}{c}{(11)}\\
				&\multicolumn{1}{c}{Aggregate}&\multicolumn{1}{c}{Food}&\multicolumn{1}{c}{Alcohol \& Tobacco}&\multicolumn{1}{c}{Clothing}&\multicolumn{1}{c}{Recreation}&\multicolumn{1}{c}{Education}&\multicolumn{1}{c}{Healthcare}&\multicolumn{1}{c}{Housing \& Utilities }&\multicolumn{1}{c}{Transportation}&\multicolumn{1}{c}{Communication}&\multicolumn{1}{c}{Hotels \& Restaurants}\\
				
					\hline 
				\midrule
				\multicolumn{12}{c}{\textbf{\emph{Panel 1: Male, Female, and Joint}}} \\ 
				\midrule 
					\hspace{0.1cm} \textbf{Section A: Non-Matrilineal} &&&&&&&& \\ 
					\midrule 
		\hspace{0.1cm} Predicted change in male income&       5.021        &       5.847        &      -1.062        &     -14.590        &       8.385        &       7.257        &      -9.556        &       6.327        &       5.430        &       7.792        &       8.488        \\
		&     (8.342)        &     (8.877)        &     (7.480)        &     (8.264)        &     (6.425)        &     (6.854)        &     (7.913)        &     (7.066)        &     (8.287)        &     (7.264)        &     (7.354)        \\
		\hspace{0.1cm} Predicted change in female income&       3.628\sym{**}&       3.517\sym{**}&       1.455        &       2.358        &       2.585\sym{**}&       2.106\sym{*} &       1.373        &       2.972\sym{**}&       2.216        &       2.998\sym{**}&       1.260        \\
		&     (1.256)        &     (1.337)        &     (1.127)        &     (1.245)        &     (0.968)        &     (1.032)        &     (1.192)        &     (1.064)        &     (1.248)        &     (1.094)        &     (1.108)        \\
		\hspace{0.1cm} Predicted change in joint income&       2.723\sym{**}&       2.731\sym{**}&       0.461        &       2.145\sym{**}&       0.436        &       0.840        &       1.705\sym{*} &       1.707\sym{**}&       0.835        &       0.260        &       0.894        \\
		&     (0.718)        &     (0.764)        &     (0.644)        &     (0.712)        &     (0.553)        &     (0.590)        &     (0.681)        &     (0.608)        &     (0.713)        &     (0.625)        &     (0.633)        \\
					\midrule
					\hspace{0.1cm}  Overidentification Wald-Test&\multicolumn{1}{c}{}         
					&\multicolumn{1}{c}{3.13}         
					&\multicolumn{1}{c}{36.08}         
					&\multicolumn{1}{c}{139.32}         
					&\multicolumn{1}{c}{26.02}         
					&\multicolumn{1}{c}{30.79}         
					&\multicolumn{1}{c}{123.07}         
					&\multicolumn{1}{c}{8.47} 
					&\multicolumn{1}{c}{13.03} 
					&\multicolumn{1}{c}{25.86}  
					&\multicolumn{1}{c}{29.46}            \\
					&\multicolumn{1}{c}{}         
					&\multicolumn{1}{c}{(0.373)}         
					&\multicolumn{1}{c}{(0.000)}         
					&\multicolumn{1}{c}{(0.000)}         
					&\multicolumn{1}{c}{(0.000)}         
					&\multicolumn{1}{c}{(0.000)}         
					&\multicolumn{1}{c}{(0.000)}         
					&\multicolumn{1}{c}{(0.037)}  
					&\multicolumn{1}{c}{(0.005)}  
					&\multicolumn{1}{c}{(0.000)}     
					&\multicolumn{1}{c}{(0.000)}               \\
				Observations        &\multicolumn{1}{c}{269}         &\multicolumn{1}{c}{269}         &\multicolumn{1}{c}{269}         &\multicolumn{1}{c}{269}         &\multicolumn{1}{c}{269}         &\multicolumn{1}{c}{269}         &\multicolumn{1}{c}{269}         &\multicolumn{1}{c}{269}         &\multicolumn{1}{c}{269}         &\multicolumn{1}{c}{269}         &\multicolumn{1}{c}{269}         \\
				\(R^{2}\)          &\multicolumn{1}{c}{0.213}        &\multicolumn{1}{c}{0.199}        &\multicolumn{1}{c}{0.045}        &\multicolumn{1}{c}{0.113}        &\multicolumn{1}{c}{0.084}        &\multicolumn{1}{c}{0.076}        &\multicolumn{1}{c}{0.089}        &\multicolumn{1}{c}{0.229}        &\multicolumn{1}{c}{0.063}        &\multicolumn{1}{c}{0.058}        &\multicolumn{1}{c}{0.099}        \\
					\midrule
					\midrule 
					\hspace{0.1cm} \textbf{Section B: Matrilineal} &&&&&&&& \\ 
					\midrule 
	\hspace{0.1cm} Predicted change in male income&       1.644\sym{*}  &       1.382         &      -0.408         &      -1.191         &      -0.154         &      -0.617         &      -0.252         &       1.441\sym{*}  &      -0.228         &      -0.539         &       1.048         \\
	&     (0.966)         &     (1.014)         &     (0.936)         &     (1.045)         &     (0.732)         &     (0.765)         &     (0.990)         &     (0.760)         &     (0.948)         &     (0.846)         &     (0.875)         \\
	\hspace{0.1cm} Predicted change in female income&       5.276\sym{***}&       5.428\sym{***}&       2.545         &       3.711\sym{**} &       4.026\sym{***}&       4.450\sym{***}&       4.967\sym{***}&       3.340\sym{**} &       2.292         &       0.899         &      -0.769         \\
	&     (1.685)         &     (1.767)         &     (1.632)         &     (1.822)         &     (1.277)         &     (1.334)         &     (1.726)         &     (1.326)         &     (1.652)         &     (1.475)         &     (1.524)         \\
	\hspace{0.1cm} Predicted change in joint income&       0.655         &       0.681         &       0.778         &       1.024         &      -0.572         &      -0.153         &      -0.502         &       0.426         &       0.941         &       0.838         &       0.775         \\
	&     (0.638)         &     (0.669)         &     (0.618)         &     (0.690)         &     (0.483)         &     (0.505)         &     (0.653)         &     (0.502)         &     (0.625)         &     (0.558)         &     (0.577)         \\
				\midrule
					\hspace{0.1cm}  Overidentification Wald-Test&\multicolumn{1}{c}{}         
					&\multicolumn{1}{c}{4.19}         
					&\multicolumn{1}{c}{186.88}         
					&\multicolumn{1}{c}{490.29}         
					&\multicolumn{1}{c}{998.09}         
					&\multicolumn{1}{c}{539.79}      
					&\multicolumn{1}{c}{715.04}         
					&\multicolumn{1}{c}{24.26} 
					&\multicolumn{1}{c}{268.85} 
					&\multicolumn{1}{c}{262.08}  
					&\multicolumn{1}{c}{154.41}            \\
					&\multicolumn{1}{c}{}         
					&\multicolumn{1}{c}{(0.241)}         
					&\multicolumn{1}{c}{(0.000)}         
					&\multicolumn{1}{c}{(0.000)}         
					&\multicolumn{1}{c}{(0.000)}         
					&\multicolumn{1}{c}{(0.000)}      
					&\multicolumn{1}{c}{(0.000)}         
					&\multicolumn{1}{c}{(0.000)}  
					&\multicolumn{1}{c}{(0.000)}  
					&\multicolumn{1}{c}{(0.000)}     
					&\multicolumn{1}{c}{(0.000)}               \\
				Observations        &\multicolumn{1}{c}{581}         &\multicolumn{1}{c}{581}         &\multicolumn{1}{c}{581}         &\multicolumn{1}{c}{581}         &\multicolumn{1}{c}{581}         &\multicolumn{1}{c}{581}         &\multicolumn{1}{c}{581}         &\multicolumn{1}{c}{581}         &\multicolumn{1}{c}{581}         &\multicolumn{1}{c}{581}         &\multicolumn{1}{c}{581}         \\
				\(R^{2}\)              &\multicolumn{1}{c}{0.224}         &\multicolumn{1}{c}{0.200}         &\multicolumn{1}{c}{0.080}         &\multicolumn{1}{c}{0.076}         &\multicolumn{1}{c}{0.056}         &\multicolumn{1}{c}{0.076}         &\multicolumn{1}{c}{0.053}         &\multicolumn{1}{c}{0.197}         &\multicolumn{1}{c}{0.068}         &\multicolumn{1}{c}{0.028}         &\multicolumn{1}{c}{0.054}         \\

				\hline 
		\midrule
		\multicolumn{12}{c}{\textbf{\emph{Panel 2: Male and Female}}} \\ 
		\midrule 
		\hspace{0.1cm} \textbf{Section A: Non-Matrilineal} &&&&&&&& \\ 
		\midrule 
\hspace{0.1cm} Predicted change in male income&       2.061\sym{**}&       2.110\sym{**}&       0.228        &       0.749\sym{*} &       0.626\sym{*} &       0.876\sym{**}&       0.711\sym{*} &       1.421\sym{**}&       0.781\sym{*} &       0.457        &       1.003\sym{**}\\
&     (0.370)        &     (0.394)        &     (0.332)        &     (0.366)        &     (0.285)        &     (0.304)        &     (0.351)        &     (0.313)        &     (0.367)        &     (0.322)        &     (0.326)        \\
\hspace{0.1cm} Predicted change in female income&       2.687\sym{**}&       2.496\sym{*} &       1.464        &       3.451\sym{**}&       1.643\sym{*} &       1.232        &       2.053\sym{*} &       2.054\sym{*} &       1.520        &       2.128\sym{*} &       0.278        \\
&     (1.031)        &     (1.097)        &     (0.924)        &     (1.021)        &     (0.794)        &     (0.847)        &     (0.978)        &     (0.873)        &     (1.024)        &     (0.898)        &     (0.909)        \\
		\midrule
		\hspace{0.1cm}  Overidentification Wald-Test&\multicolumn{1}{c}{}         
		&\multicolumn{1}{c}{2.62}         
		&\multicolumn{1}{c}{22.15}         
		&\multicolumn{1}{c}{10.23}         
		&\multicolumn{1}{c}{18.03}         
		&\multicolumn{1}{c}{26.53}         
		&\multicolumn{1}{c}{12.75}         
		&\multicolumn{1}{c}{7.25} 
		&\multicolumn{1}{c}{12.69} 
		&\multicolumn{1}{c}{18.79}  
		&\multicolumn{1}{c}{22.15}            \\
		&\multicolumn{1}{c}{}         
		&\multicolumn{1}{c}{(0.269)}         
		&\multicolumn{1}{c}{(0.000)}         
		&\multicolumn{1}{c}{(0.000)}         
		&\multicolumn{1}{c}{(0.000)}         
		&\multicolumn{1}{c}{(0.000)}         
		&\multicolumn{1}{c}{(0.000)}         
		&\multicolumn{1}{c}{(0.027)}  
		&\multicolumn{1}{c}{(0.002)}  
		&\multicolumn{1}{c}{(0.000)}     
		&\multicolumn{1}{c}{(0.000)}               \\
		Observations        &\multicolumn{1}{c}{269}         &\multicolumn{1}{c}{269}         &\multicolumn{1}{c}{269}         &\multicolumn{1}{c}{269}         &\multicolumn{1}{c}{269}         &\multicolumn{1}{c}{269}         &\multicolumn{1}{c}{269}         &\multicolumn{1}{c}{269}         &\multicolumn{1}{c}{269}         &\multicolumn{1}{c}{269}         &\multicolumn{1}{c}{269}         \\
		\(R^{2}\)               &\multicolumn{1}{c}{0.213}        &\multicolumn{1}{c}{0.199}        &\multicolumn{1}{c}{0.045}        &\multicolumn{1}{c}{0.113}        &\multicolumn{1}{c}{0.084}        &\multicolumn{1}{c}{0.076}        &\multicolumn{1}{c}{0.089}        &\multicolumn{1}{c}{0.229}        &\multicolumn{1}{c}{0.063}        &\multicolumn{1}{c}{0.058}        &\multicolumn{1}{c}{0.099}        \\
		\midrule
		\midrule 
		\hspace{0.1cm} \textbf{Section B: Matrilineal} &&&&&&&& \\ 
		\midrule
\hspace{0.1cm} Predicted change in male income&       0.689        &       0.397        &      -0.927        &      -1.939        &      -0.773        &      -1.356        &      -1.039        &       0.838        &      -0.721        &      -0.787        &       1.100        \\
&     (1.129)        &     (1.184)        &     (1.094)        &     (1.221)        &     (0.856)        &     (0.894)        &     (1.157)        &     (0.888)        &     (1.107)        &     (0.989)        &     (1.022)        \\
\hspace{0.1cm} Predicted change in female income&       5.745        &       6.539        &       6.226        &      10.002\sym{**}&       4.678        &       7.171\sym{**}&       6.368        &       2.869        &       5.888        &       4.292        &      -1.361        \\
&     (3.454)        &     (3.623)        &     (3.346)        &     (3.736)        &     (2.618)        &     (2.734)        &     (3.538)        &     (2.718)        &     (3.386)        &     (3.025)        &     (3.125)        \\
		\midrule 
		\hspace{0.1cm}  Overidentification Wald-Test&\multicolumn{1}{c}{}         
		&\multicolumn{1}{c}{3.34}         
		&\multicolumn{1}{c}{155.49}         
		&\multicolumn{1}{c}{404.64}         
		&\multicolumn{1}{c}{216.83}         
		&\multicolumn{1}{c}{233.04}      
		&\multicolumn{1}{c}{96.13}         
		&\multicolumn{1}{c}{18.46} 
		&\multicolumn{1}{c}{177.89} 
		&\multicolumn{1}{c}{169.11}  
		&\multicolumn{1}{c}{71.06}            \\
		&\multicolumn{1}{c}{}         
		&\multicolumn{1}{c}{(0.189)}         
		&\multicolumn{1}{c}{(0.000)}         
		&\multicolumn{1}{c}{(0.000)}         
		&\multicolumn{1}{c}{(0.014)}         
		&\multicolumn{1}{c}{(0.000)}      
		&\multicolumn{1}{c}{(0.000)}         
		&\multicolumn{1}{c}{(0.000)}  
		&\multicolumn{1}{c}{(0.000)}  
		&\multicolumn{1}{c}{(0.000)}     
		&\multicolumn{1}{c}{(0.000)}               \\
		Observations        &\multicolumn{1}{c}{581}         &\multicolumn{1}{c}{581}         &\multicolumn{1}{c}{581}         &\multicolumn{1}{c}{581}         &\multicolumn{1}{c}{581}         &\multicolumn{1}{c}{581}         &\multicolumn{1}{c}{581}         &\multicolumn{1}{c}{581}         &\multicolumn{1}{c}{581}         &\multicolumn{1}{c}{581}         &\multicolumn{1}{c}{581}         \\
		\(R^{2}\)         &\multicolumn{1}{c}{0.224}        &\multicolumn{1}{c}{0.200}        &\multicolumn{1}{c}{0.080}        &\multicolumn{1}{c}{0.076}        &\multicolumn{1}{c}{0.056}        &\multicolumn{1}{c}{0.076}        &\multicolumn{1}{c}{0.053}        &\multicolumn{1}{c}{0.197}        &\multicolumn{1}{c}{0.068}        &\multicolumn{1}{c}{0.028}        &\multicolumn{1}{c}{0.054}        \\

					\hline 
					\hline \\[-1.8ex] 
					\multicolumn{12}{p{\linewidth}}{\footnotesize  \textit{Note}: N = 850. The table presents coefficients of the difference in log consumption of each item on the difference in predicted log income, as from Table~\ref{tab:e1firststagemfmfj}. Regressions include agro-ecological zone indicators. Fully robust standard errors clustered at the household are in parentheses  (\sym{*} \(p<0.05\), \sym{**} \(p<0.01\)) below the predicted change variables. Below the Wald test in parentheses is the probability $> \chi^2$.} \\
			\end{tabular}}
			\setbox0=\hbox{\contents}
			\setlength{\linewidth}{\wd0-2\tabcolsep-.25em}
			\contents}
	\end{table}
\end{landscape}

\begin{landscape}	
	\begin{table}[htbp]\centering
		\def\sym#1{\ifmmode^{#1}\else\(^{#1}\)\fi}
		\caption{Restricted Overidentification Tests - Log of Consumption (Female-Headed) \label{tab:e1s2estimatesfem}}
		\resizebox{9in}{!}
		{\setlength{\linewidth}{.1cm}\newcommand{\contents}
			{\begin{tabular}{l*{12}{D{.}{.}{-1}}}
					\\[-1.8ex]\hline 
					\hline \\[-1.8ex]
					&\multicolumn{1}{c}{(1)}&\multicolumn{1}{c}{(2)}&\multicolumn{1}{c}{(3)}&\multicolumn{1}{c}{(4)}&\multicolumn{1}{c}{(5)}&\multicolumn{1}{c}{(6)}&\multicolumn{1}{c}{(7)}&\multicolumn{1}{c}{(8)}&\multicolumn{1}{c}{(9)}&\multicolumn{1}{c}{(10)}&\multicolumn{1}{c}{(11)}\\
					&\multicolumn{1}{c}{Aggregate}&\multicolumn{1}{c}{Food}&\multicolumn{1}{c}{Alcohol \& Tobacco}&\multicolumn{1}{c}{Clothing}&\multicolumn{1}{c}{Recreation}&\multicolumn{1}{c}{Education}&\multicolumn{1}{c}{Healthcare}&\multicolumn{1}{c}{Housing \& Utilities }&\multicolumn{1}{c}{Transportation}&\multicolumn{1}{c}{Communication}&\multicolumn{1}{c}{Hotels \& Restaurants}\\
					
					\hline 
					\midrule
					\multicolumn{12}{c}{\textbf{\emph{Panel 1: Male, Female, and Joint}}} \\ 
					\midrule 
					
					\hspace{0.1cm} \textbf{Section A: Non-Female-Headed} &&&&&&&& \\ 
					\midrule 
				\hspace{0.1cm} Predicted change in male income&       1.608        &       1.460        &      -2.443        &      -5.472\sym{**}&      -0.536        &       0.772        &      -1.875        &      -0.619        &      -2.820        &      -1.955        &       1.877        \\
				&     (1.747)        &     (1.857)        &     (1.694)        &     (1.871)        &     (1.398)        &     (1.384)        &     (1.769)        &     (1.440)        &     (1.772)        &     (1.614)        &     (1.607)        \\
				\hspace{0.1cm} Predicted change in female income&      13.805\sym{**}&      14.604\sym{**}&       7.605        &      10.979\sym{*} &      16.898\sym{**}&       9.638\sym{*} &      10.398\sym{*} &      15.327\sym{**}&      13.540\sym{**}&       9.234\sym{*} &       0.338        \\
				&     (4.819)        &     (5.122)        &     (4.673)        &     (5.161)        &     (3.857)        &     (3.818)        &     (4.879)        &     (3.973)        &     (4.887)        &     (4.452)        &     (4.432)        \\
				\hspace{0.1cm} Predicted change in joint income&       0.857        &       0.855        &       1.049        &       2.380\sym{**}&      -0.603        &      -0.373        &       0.596        &       0.582        &       0.946        &       0.512        &       0.232        \\
				&     (0.798)        &     (0.848)        &     (0.774)        &     (0.855)        &     (0.639)        &     (0.632)        &     (0.808)        &     (0.658)        &     (0.809)        &     (0.737)        &     (0.734)        \\
				\midrule
					\hspace{0.1cm}  Overidentification Wald-Test&\multicolumn{1}{c}{}         
					&\multicolumn{1}{c}{0.49}         
					&\multicolumn{1}{c}{731.10}         
					&\multicolumn{1}{c}{332.45}         
					&\multicolumn{1}{c}{374.67}         
					&\multicolumn{1}{c}{88.83}         
					&\multicolumn{1}{c}{357.82}         
					&\multicolumn{1}{c}{18.67} 
					&\multicolumn{1}{c}{563.16} 
					&\multicolumn{1}{c}{878.65}  
					&\multicolumn{1}{c}{48.80}            \\
					&\multicolumn{1}{c}{}         
					&\multicolumn{1}{c}{(0.921)}         
					&\multicolumn{1}{c}{(0.000)}         
					&\multicolumn{1}{c}{(0.000)}         
					&\multicolumn{1}{c}{(0.000)}         
					&\multicolumn{1}{c}{(0.000)}         
					&\multicolumn{1}{c}{(0.000)}         
					&\multicolumn{1}{c}{(0.000)}  
					&\multicolumn{1}{c}{(0.000)}  
					&\multicolumn{1}{c}{(0.000)}     
					&\multicolumn{1}{c}{(0.000)}               \\
			Observations        &\multicolumn{1}{c}{677}         &\multicolumn{1}{c}{677}         &\multicolumn{1}{c}{677}         &\multicolumn{1}{c}{677}         &\multicolumn{1}{c}{677}         &\multicolumn{1}{c}{677}         &\multicolumn{1}{c}{677}         &\multicolumn{1}{c}{677}         &\multicolumn{1}{c}{677}         &\multicolumn{1}{c}{677}         &\multicolumn{1}{c}{677}         \\
			\(R^{2}\)           &\multicolumn{1}{c}{0.220}        &\multicolumn{1}{c}{0.205}        &\multicolumn{1}{c}{0.061}        &\multicolumn{1}{c}{0.081}        &\multicolumn{1}{c}{0.081}        &\multicolumn{1}{c}{0.062}        &\multicolumn{1}{c}{0.053}        &\multicolumn{1}{c}{0.210}        &\multicolumn{1}{c}{0.075}        &\multicolumn{1}{c}{0.034}        &\multicolumn{1}{c}{0.048}        \\
					\midrule
					\midrule
					\hspace{0.1cm} \textbf{Section B: Female-Headed} &&&&&&&& \\ 
					\midrule
\hspace{0.1cm} Predicted change in male income&     -17.254        &     -13.829        &     -20.866        &      -5.289        &       3.530        &     -15.054        &     -20.371        &     -34.773\sym{**}&     -22.431        &       1.491        &     -13.172        \\
&    (16.195)        &    (16.332)        &    (14.103)        &    (16.496)        &     (8.961)        &    (13.536)        &    (15.677)        &    (12.219)        &    (13.935)        &    (10.556)        &    (13.346)        \\
\hspace{0.1cm} Predicted change in female income&       1.551\sym{*} &       1.559\sym{*} &       0.094        &       0.926        &       0.026        &       1.195\sym{*} &       0.475        &      -0.306        &      -0.618        &       0.722        &      -0.016        \\
&     (0.711)        &     (0.717)        &     (0.619)        &     (0.724)        &     (0.393)        &     (0.594)        &     (0.688)        &     (0.536)        &     (0.612)        &     (0.463)        &     (0.586)        \\
\hspace{0.1cm} Predicted change in joint income&       3.114\sym{*} &       3.109\sym{*} &       0.728        &       1.337        &       0.411        &       0.700        &       1.600        &      -0.077        &       0.646        &       0.242        &      -0.177        \\
&     (1.210)        &     (1.221)        &     (1.054)        &     (1.233)        &     (0.670)        &     (1.012)        &     (1.172)        &     (0.913)        &     (1.041)        &     (0.789)        &     (0.997)        \\
					\midrule 
					\hspace{0.1cm}  Overidentification Wald-Test&\multicolumn{1}{c}{}         
					&\multicolumn{1}{c}{1.95}         
					&\multicolumn{1}{c}{337.29}         
					&\multicolumn{1}{c}{60.22}         
					&\multicolumn{1}{c}{411.48}         
					&\multicolumn{1}{c}{27.42}      
					&\multicolumn{1}{c}{203.48}         
					&\multicolumn{1}{c}{633.58} 
					&\multicolumn{1}{c}{785.60} 
					&\multicolumn{1}{c}{606.50}  
					&\multicolumn{1}{c}{371.41}            \\
					&\multicolumn{1}{c}{}         
					&\multicolumn{1}{c}{(0.583)}         
					&\multicolumn{1}{c}{(0.000)}         
					&\multicolumn{1}{c}{(0.000)}         
					&\multicolumn{1}{c}{(0.000)}         
					&\multicolumn{1}{c}{(0.000)}      
					&\multicolumn{1}{c}{(0.000)}         
					&\multicolumn{1}{c}{(0.000)}  
					&\multicolumn{1}{c}{(0.000)}  
					&\multicolumn{1}{c}{(0.000)}     
					&\multicolumn{1}{c}{(0.139)}               \\
Observations        &\multicolumn{1}{c}{173}         &\multicolumn{1}{c}{173}         &\multicolumn{1}{c}{173}         &\multicolumn{1}{c}{173}         &\multicolumn{1}{c}{173}         &\multicolumn{1}{c}{173}         &\multicolumn{1}{c}{173}         &\multicolumn{1}{c}{173}         &\multicolumn{1}{c}{173}         &\multicolumn{1}{c}{173}         &\multicolumn{1}{c}{173}         \\
\(R^{2}\)         &\multicolumn{1}{c}{0.252}        &\multicolumn{1}{c}{0.225}        &\multicolumn{1}{c}{0.093}        &\multicolumn{1}{c}{0.096}        &\multicolumn{1}{c}{0.076}        &\multicolumn{1}{c}{0.158}        &\multicolumn{1}{c}{0.118}        &\multicolumn{1}{c}{0.212}        &\multicolumn{1}{c}{0.046}        &\multicolumn{1}{c}{0.050}        &\multicolumn{1}{c}{0.146}        \\

					\hline 
					\midrule
					\multicolumn{12}{c}{\textbf{\emph{Panel 2: Male and Female}}} \\ 
					\midrule 
					\hspace{0.1cm} \textbf{Section A: Non-Female Headed} &&&&&&&& \\ 
					\midrule 
		\hspace{0.1cm} Predicted change in male income&       0.861        &       0.850        &       0.823        &       1.882\sym{*} &      -0.621        &      -0.320        &       0.429        &       0.478        &       0.695        &       0.349        &       0.320        \\
		&     (0.684)        &     (0.727)        &     (0.663)        &     (0.733)        &     (0.548)        &     (0.542)        &     (0.693)        &     (0.564)        &     (0.694)        &     (0.632)        &     (0.629)        \\
		\hspace{0.1cm} Predicted change in female income&      14.048        &      14.539        &      -0.561        &      -7.248        &      17.237\sym{**}&      12.041\sym{*} &       4.647        &      11.970        &       4.795        &       3.607        &       3.330        \\
		&     (7.603)        &     (8.080)        &     (7.371)        &     (8.142)        &     (6.085)        &     (6.023)        &     (7.697)        &     (6.267)        &     (7.709)        &     (7.023)        &     (6.992)        \\
					\midrule
					\hspace{0.1cm}  Overidentification Wald-Test&\multicolumn{1}{c}{}         
					&\multicolumn{1}{c}{0.26}         
					&\multicolumn{1}{c}{60.65}         
					&\multicolumn{1}{c}{171.26}         
					&\multicolumn{1}{c}{136.08}         
					&\multicolumn{1}{c}{87.19}         
					&\multicolumn{1}{c}{33.31}         
					&\multicolumn{1}{c}{15.85} 
					&\multicolumn{1}{c}{26.58} 
					&\multicolumn{1}{c}{47.56}  
					&\multicolumn{1}{c}{48.10}            \\
					&\multicolumn{1}{c}{}         
					&\multicolumn{1}{c}{(0.880)}         
					&\multicolumn{1}{c}{(0.000)}         
					&\multicolumn{1}{c}{(0.000)}         
					&\multicolumn{1}{c}{(0.000)}         
					&\multicolumn{1}{c}{(0.000)}         
					&\multicolumn{1}{c}{(0.000)}         
					&\multicolumn{1}{c}{(0.000)}  
					&\multicolumn{1}{c}{(0.000)}  
					&\multicolumn{1}{c}{(0.000)}     
					&\multicolumn{1}{c}{(0.000)}               \\
					Observations           &\multicolumn{1}{c}{677}         &\multicolumn{1}{c}{677}         &\multicolumn{1}{c}{677}         &\multicolumn{1}{c}{677}         &\multicolumn{1}{c}{677}         &\multicolumn{1}{c}{677}         &\multicolumn{1}{c}{677}         &\multicolumn{1}{c}{677}         &\multicolumn{1}{c}{677}         &\multicolumn{1}{c}{677}         &\multicolumn{1}{c}{677}         \\
					\(R^{2}\)           &\multicolumn{1}{c}{0.220}        &\multicolumn{1}{c}{0.205}        &\multicolumn{1}{c}{0.061}        &\multicolumn{1}{c}{0.081}        &\multicolumn{1}{c}{0.081}        &\multicolumn{1}{c}{0.062}        &\multicolumn{1}{c}{0.053}        &\multicolumn{1}{c}{0.210}        &\multicolumn{1}{c}{0.075}        &\multicolumn{1}{c}{0.034}        &\multicolumn{1}{c}{0.048}        \\
					\midrule
					\midrule 
					\hspace{0.1cm} \textbf{Section B: Female-Headed} &&&&&&&& \\ 
					\midrule
			\hspace{0.1cm} Predicted change in male income&     -19.360        &     -15.905        &     -21.994        &      -5.577        &       2.829        &     -13.526        &     -22.215        &     -35.353\sym{**}&     -25.091        &       2.750        &     -12.881        \\
			&    (16.783)        &    (16.925)        &    (14.615)        &    (17.095)        &     (9.286)        &    (14.027)        &    (16.246)        &    (12.663)        &    (14.440)        &    (10.939)        &    (13.830)        \\
			\hspace{0.1cm} Predicted change in female income&       1.524\sym{*} &       1.533\sym{*} &       0.083        &       0.919        &       0.019        &       1.203\sym{*} &       0.455        &      -0.310        &      -0.640        &       0.731        &      -0.013        \\
			&     (0.715)        &     (0.721)        &     (0.623)        &     (0.729)        &     (0.396)        &     (0.598)        &     (0.693)        &     (0.540)        &     (0.616)        &     (0.466)        &     (0.590)        \\
					\midrule 
					\hspace{0.1cm}  Overidentification Wald-Test&\multicolumn{1}{c}{}         
					&\multicolumn{1}{c}{1.60}         
					&\multicolumn{1}{c}{249.64}         
					&\multicolumn{1}{c}{39.76}         
					&\multicolumn{1}{c}{356.35}         
					&\multicolumn{1}{c}{13.51}      
					&\multicolumn{1}{c}{126.35}         
					&\multicolumn{1}{c}{536.35} 
					&\multicolumn{1}{c}{487.23} 
					&\multicolumn{1}{c}{119.47}  
					&\multicolumn{1}{c}{296.85}            \\
					&\multicolumn{1}{c}{}         
					&\multicolumn{1}{c}{(0.448)}         
					&\multicolumn{1}{c}{(0.000)}         
					&\multicolumn{1}{c}{(0.000)}         
					&\multicolumn{1}{c}{(0.000)}         
					&\multicolumn{1}{c}{(0.001)}      
					&\multicolumn{1}{c}{(0.000)}         
					&\multicolumn{1}{c}{(0.000)}  
					&\multicolumn{1}{c}{(0.000)}  
					&\multicolumn{1}{c}{(0.000)}     
					&\multicolumn{1}{c}{(0.000)}               \\
					Observations        &\multicolumn{1}{c}{173}         &\multicolumn{1}{c}{173}         &\multicolumn{1}{c}{173}         &\multicolumn{1}{c}{173}         &\multicolumn{1}{c}{173}         &\multicolumn{1}{c}{173}         &\multicolumn{1}{c}{173}         &\multicolumn{1}{c}{173}         &\multicolumn{1}{c}{173}         &\multicolumn{1}{c}{173}         &\multicolumn{1}{c}{173}         \\
					\(R^{2}\)                  &\multicolumn{1}{c}{0.252}        &\multicolumn{1}{c}{0.225}        &\multicolumn{1}{c}{0.093}        &\multicolumn{1}{c}{0.096}        &\multicolumn{1}{c}{0.076}        &\multicolumn{1}{c}{0.158}        &\multicolumn{1}{c}{0.118}        &\multicolumn{1}{c}{0.212}        &\multicolumn{1}{c}{0.046}        &\multicolumn{1}{c}{0.050}        &\multicolumn{1}{c}{0.146}        \\
				
					\hline 
					\hline \\[-1.8ex] 
					\multicolumn{12}{p{\linewidth}}{\footnotesize  \textit{Note}: N = 850. The table presents coefficients of the difference in log consumption of each item on the difference in predicted log income, as from Table~\ref{tab:e1firststagemfmfj}. Regressions include agro-ecological zone indicators. Fully robust standard errors clustered at the household are in parentheses  (\sym{*} \(p<0.05\), \sym{**} \(p<0.01\)) below the predicted change variables. Below the Wald test in parentheses is the probability $> \chi^2$.} \\
			\end{tabular}}
			\setbox0=\hbox{\contents}
			\setlength{\linewidth}{\wd0-2\tabcolsep-.25em}
			\contents}
	\end{table}
\end{landscape}

\newpage

\appendix 

\section{Appendix}

\setcounter{table}{0}
\renewcommand\thetable{\Alph{section}.\arabic{table}}

\onehalfspacing

\subsection{Prices and Price Stability}\label{sec:e1pricesec}

An essential assumption in the theoretical and empirical models is that changes in consumption are driven by only changes in rainfall. This assumption implies that there is no indirect link between rainfall and consumption. An indirect link might exist if changes in rainfall induce changes in price, which impact consumption and thus household expenditures. In this section I briefly discuss some of the relevant literature which supports this assumption.

\cite{Timmer00fp} suggests that price stability has an essential role in the process of structural transformation and policies which work to stabilize prices are common around the world. Additional evidence in this area is supported by \cite{Foxetal11, Chavasetal05}. Price stability observed is often the result of government and development agency policy efforts throughout Southern Africa \citep{Jaynetal06, dHoteletal12}. Efforts towards coordination, transparency, and consultation between stakeholders have improved price stability in recent years. Much of this achievement has come from the compiling of strategic grain reserves, as modeled in \cite{Tranetal15} and \cite{MasonandMyers13}. 

Several studies, from Southern Africa generally and Malawi in particular, have explored the existence of price stability in cases where changes would be expected to shift prices. \cite{JRGetal13} considers the impact of fertilizer subsidy programs on maize price stability in Zambia and Malawi: they find that even doubling the subsidy results in only small price changes. The conclusions of this work are supported by \cite{Denningetal09} who determine that input subsidies and maize surpluses improved price stability in Malawi. 

Price stability is a reasonable assumption across much of Southern Africa, including Malawi, due to the prevalence of such programs and efforts to stabilize prices throughout the region. These findings and relevant literature support the assumption that changes in consumption are driven by only changes in rainfall experienced.  

\subsection{Additional Tests and Analyses}\label{sec:unrestricoverid}
	
Table~\ref{tab:e1unconestimates} presents the unconstrained estimates of the relationship between expenditure and rainfall. For each regression, nine rainfall variables, are included, as well as location indicators. These results are not disaggregated by gender and hence cannot address the potentially gendered nature of income earning and expenditure. 
	

\newpage
\begin{landscape}

	\begin{table}[h]\centering
		\def\sym#1{\ifmmode^{#1}\else\(^{#1}\)\fi}
		\caption{Unrestricted Overidentification Tests \label{tab:e1unconestimates}}
		\resizebox{9in}{!}
	{\setlength{\linewidth}{.1cm}\newcommand{\contents}
		{\begin{tabular}{l*{12}{D{.}{.}{-1}}}
				\\[-1.8ex]\hline 
				\hline \\[-1.8ex]
				&\multicolumn{1}{c}{(1)}&\multicolumn{1}{c}{(2)}&\multicolumn{1}{c}{(3)}&\multicolumn{1}{c}{(4)}&\multicolumn{1}{c}{(5)}&\multicolumn{1}{c}{(6)}&\multicolumn{1}{c}{(7)}&\multicolumn{1}{c}{(8)}&\multicolumn{1}{c}{(9)}&\multicolumn{1}{c}{(10)}&\multicolumn{1}{c}{(11)}\\
				&\multicolumn{1}{c}{Aggregate}&\multicolumn{1}{c}{Food}&\multicolumn{1}{c}{Alcohol \& Tobacco}&\multicolumn{1}{c}{Clothing}&\multicolumn{1}{c}{Recreation}&\multicolumn{1}{c}{Education}&\multicolumn{1}{c}{Healthcare}&\multicolumn{1}{c}{Housing \& Utilities }&\multicolumn{1}{c}{Transportation}&\multicolumn{1}{c}{Communication}&\multicolumn{1}{c}{Hotels \& Restaurants}\\
				\hline 
				\midrule 
				\hspace{0.1cm} Total seasonal rainfall    &     0.00447\sym{**}&     0.00446\sym{**}&     0.00167\sym{**}&     0.00249\sym{**}&     0.00154\sym{**}&     0.00189\sym{**}&     0.00196\sym{**}&     0.00313\sym{**}&     0.00208\sym{**}&     0.00127\sym{**}&     0.00132\sym{**}\\
				&   (0.00035)        &   (0.00034)        &   (0.00030)        &   (0.00031)        &   (0.00025)        &   (0.00025)        &   (0.00031)        &   (0.00026)        &   (0.00030)        &   (0.00028)        &   (0.00028)        \\
				\midrule

				\hspace{0.1cm}  \(R^{2}\)             &\multicolumn{1}{c}{0.214}        &\multicolumn{1}{c}{0.196}        &\multicolumn{1}{c}{0.064}        &\multicolumn{1}{c}{0.080}        &\multicolumn{1}{c}{0.060}        &\multicolumn{1}{c}{0.075}        &\multicolumn{1}{c}{0.061}        &\multicolumn{1}{c}{0.201}        &\multicolumn{1}{c}{0.060}        &\multicolumn{1}{c}{0.031}        &\multicolumn{1}{c}{0.056}        \\

				\hline 
				\hline \\[-1.8ex] 
				\multicolumn{12}{p{\linewidth}}{\footnotesize  \textit{Note}: N = 850. Regressions include agro-ecological zone indicators. Fully robust standard errors clustered at the household are in parentheses  (\sym{*} \(p<0.05\), \sym{**} \(p<0.01\)).} \\
		\end{tabular}}
		\setbox0=\hbox{\contents}
		\setlength{\linewidth}{\wd0-2\tabcolsep-.25em}
		\contents}
\end{table}

\end{landscape}

\end{document}